\documentclass[12pt]{article}
\usepackage{hyperref}
\usepackage[top=2cm,bottom=3cm,left=2cm,right=2cm]{geometry}
\usepackage{graphicx}
\usepackage{amsmath}
\usepackage{amssymb}

\numberwithin{equation}{section}
\numberwithin{figure}{section}

\def\eq#1{(\ref{eq:#1})}

\newcommand{\Tr}{\mathop{\rm Tr}\nolimits}
\def\d{\partial}

\def\H{\mathcal{H}}
\def\P{\mathcal{P}}
\def\Ppre{\mathcal{P}_\mathrm{pre}}
\def\L{\mathcal{L}}
\def\vol{\mathrm{vol}}

\begin{document}

\begin{titlepage}
	
	\rightline{\tt MIT-CTP-5880}
	\hfill \today
	\begin{center}
		\vskip 2cm
		
		{\Large \bf {Covariant phase space and $L_\infty$ algebras}
		}
		
		\vskip 0.5cm
		
		\vskip 1.0cm
		{\large {Vin\'{\i}cius Bernardes$^{1}$, Theodore Erler$^{1,2}$, and Atakan Hilmi F{\i}rat$^{3,4}$ }}
		
		\vskip 0.5cm
		
		{\em  \hskip -.1truecm
			$^{1}$
			CEICO, FZU - Institute of Physics of the Czech Academy of Sciences \\
			No Slovance 2, 182 21, Prague 8, Czech Republic
			\\
			\vskip 0.5cm
			$^{2}$
			Kavli Institute for Theoretical Physics, 
			University of California \\
			Santa Barbara, CA 93106-4030 
			\\
			\vskip 0.5cm
			$^{3}$
			Center for Theoretical Physics - a Leinweber Institute,
			Massachusetts Institute of Technology\\
			Cambridge, MA 02139, USA
			\\
			\vskip 0.5cm
			$^{4}$
			Center for Quantum Mathematics and Physics (QMAP),
			Department of Physics \& Astronomy, \\
			University of California, Davis, CA 95616, USA
			\\
			\vskip 0.5cm
			\tt \href{mailto:viniciusbernsilva@gmail.com}{viniciusbernsilva@gmail.com},
		\href{mailto:tchovi@gmail.com}{tchovi@gmail.com},
		 \href{mailto:ahfirat@ucdavis.edu}{ahfirat@ucdavis.edu} \vskip 5pt }
		
		\vskip 2.0cm
		{\bf Abstract}
		
	\end{center}
	\vskip 0.25cm
	\noindent
	\begin{narrower}
		\baselineskip15pt
		We propose a symplectic structure for the phase space of a generic Lagrangian field theory expressed in the framework of $L_\infty$ algebras. The symplectic structure does not require explicit knowledge of the derivative content of the Lagrangian, and therefore is applicable to nonlocal models, such as string field theory, where traditional constructions are difficult to apply. We test our proposal in a number of examples ranging from general relativity to $p$-adic string theory. 
	\end{narrower}
\end{titlepage}

\tableofcontents
\baselineskip15pt

\section{Introduction}

Most physical systems are expected to have a phase space, i.e., a collection of positions $q^i$ and momenta $p_i$ forming a manifold with symplectic structure $\delta p_i \wedge \delta q^i$. However the existence of phase space is obscure when the dynamics of the system is not local in time. First there is a problem of interpretation. In the canonical formalism, phase space is understood as the space of initial value data needed to solve the field equations. But if the field equations are not local, solutions might not be generated from data specified at a fixed time. This forces us to adopt the point of view of the {\it covariant phase space formalism}~\cite{Crnkovic,Crncovic2,Zuckerman,Lee} (see~\cite{Khavkine,Gieres,Harlow} for reviews), where phase space is understood as the space of classical solutions equated up to gauge transformation. In the covariant phase space formalism, it is not necessary to assume that solutions are parameterized through initial value data or in any other way.

However, there is a second, more concrete problem with phase space in the presence of nonlocality. There is no satisfactory way to define a symplectic structure. Ordinarily, when the Lagrangian is local, the symplectic structure is determined using either canonical or covariant phase space methods from the Lagrangian’s dependence on derivatives of the field. In principle this can be applied to a nonlocal Lagrangian if it is expressed through a derivative expansion, though there can be subtleties with the infinite derivative limit.  However, often the derivative structure of a nonlocal Lagrangian is too unwieldy to carry out this procedure in practice. 

In this paper we give a new way to compute the symplectic structure on phase space. The starting point is a  Lagrangian field theory expressed in terms of a homotopy Lie algebra, or {\it  $L_\infty$~algebra} \cite{Zwiebach,Lada,Lada2}. This is natural in the context of Batalin-Vilkovisky (BV) quantization \cite{Batalin,Henneaux}, and furnishes the ingredients to write a compact formula for the symplectic structure,
\begin{align} 
	\Omega = \frac{1}{2}\omega\big(\delta \Phi, [Q_\Phi, \sigma] \delta\Phi\big), \label{eq:Omega}
\end{align}
where
\begin{itemize}
\item $Q_\Phi$ is the kinetic operator around a solution $\Phi$,
\item $\delta$ is the exterior derivative on the space of solutions,
\item $\omega$ is the BV inner product, 
\end{itemize}
and finally $\sigma$ is an operator called the {\it sigmoid}. The sigmoid satisfies boundary conditions in the infinite past and future:
\begin{align}
	\lim_{t \to -\infty} \sigma = 0, \quad
	\lim_{t \to \infty} \sigma = 1.
\end{align}
The sigmoid's transition between past and future plays the role of a time slice. This and other ingredients will be explained in more detail below. The major advantage of this formula is that it does not require understanding how the Lagrangian depends on derivatives of the field. It applies just as easily to a nonlocal field theory as to a local one. 

The formula is related to an old proposal of Witten \cite{Witten} for defining a symplectic structure on the phase space of open string field theory \cite{Witten2}. The idea was revived in recent work of Cho, Mazel and Yin \cite{Cho}, who applied it to compute the energy of rolling tachyon solutions on unstable D-branes~\cite{Sen}. We will not discuss string field theory in the present work, though the language of $L_\infty$ algebras is very natural in this context. For us the main role of $L_\infty$ algebras is to provide a universal language for expressing the symplectic structure and demonstrating its consistency. In recent years it has been recognized that any Lagrangian field theory can be expressed in terms of an appropriate $L_\infty$ algebra \cite{Hohm,Jurco}. This follows as an extension of the universality of BV quantization in field theory. However, as we will show, the formula \eq{Omega} works to compute the symplectic structure even if the $L_\infty$ description of the theory is not explicitly available. 

A great deal of effort has been devoted to understanding the phase space of nonlocal theories down through the years. Concerning the symplectic structure, a notable approach is the formalism of \cite{Llosa,Gomis,Gomis2} which represents phase space in terms of a constrained Hamiltonian system with an auxiliary time parameter. This has been applied to $p$-adic string theory and the truncated tachyon action of open string field theory \cite{Gomis2,Heredia}. Some other more specific approaches can be found in~\cite{Barci,Woodard,Talaganis,Chang}. A major concern in studies of nonlocality is the fate of the initial value problem and higher derivative instabilities~\cite{Eliezer}. We do not address these questions here. The covariant phase space formalism refers to a hypothetical space of solutions but does not give a way to construct it, nor does it guarantee that solutions are free of higher derivative instability or other pathology.

This paper is organized as follows. Section \ref{sec:symplectic} describes the generalities of our proposal. After laying the groundwork with a short introduction to the $L_\infty$ formulation of field theory, subsection \ref{subsec:Linfty} uses this language to discuss pre-phase space, phase space, and respective complexes of differential forms which are the basis for the covariant phase space formalism. Subsection \ref{subsec:Omega} describes our proposal for the symplectic structure. We introduce the sigmoid operator, outline the conditions it must satisfy, and explain the sense that it defines a notion of time slice for the symplectic structure.  Subsection \ref{subsec:consistency} lays out a list of properties that the symplectic structure must satisfy. Perhaps the most elementary is that the symplectic structure is not zero. Understanding why this is the case requires being careful about boundary contributions when integrating by parts. We introduce the {\it tau regulator} as a means of keeping track of this. The tau regulator allows a simple proof of all remaining consistency properties of the symplectic form. Section \ref{sec:examples} illustrates the symplectic structure in examples. First we write the formula \eq{Omega} in a way that can be applied to any Lagrangian regardless of whether it has been expressed in $L_\infty$ form.  Subsections \ref{subsec:scalar} and \ref{subsec:YM} use this to evaluate the symplectic structure in scalar field theory and Yang-Mills theory. Subsection \ref{subsec:GR} computes the symplectic structure of general relativity. Following the setup of Harlow and Wu \cite{Harlow} we consider general relativity with a finite spatial boundary, and show that \eq{Omega} produces the correct symplectic structure including contributions from the boundary of the Cauchy surface. Subsection \ref{subsec:padic} evaluates the symplectic structure of $p$-adic string theory~\cite{Brekke}. Because the theory is nonlocal, the symplectic structure cannot be defined on a Cauchy surface, but has some irreducible ``fuzziness'' on the order of the string length. We test the symplectic structure by computing the energy of rolling tachyon solutions analogously to~\cite{Cho}, finding agreement with the energy formula of Moeller and Zwiebach~\cite{Moeller}. In section~\ref{sec:prospects} we conclude by listing some directions for further development.  

\subsubsection*{Conventions}

We assume that the higher order products of the $L_\infty$ algebra are graded commutative and carry grade $+1$.  The metric is taken to have mostly positive signature. 

\section{Symplectic structure}
\label{sec:symplectic}

To begin we describe the covariant phase space formalism from the point of view of an action expressed in terms of an $L_\infty$ algebra. Then we can introduce the symplectic structure and explain the mechanisms behind its consistency.

\subsection{Covariant phase space of an $L_\infty$ action}
\label{subsec:Linfty}

The $L_\infty$ approach to field theory is based on a classical action of the form
\begin{align} 
	S = - {1 \over 2} \omega (\Phi, Q \Phi) 
	- \sum_{n=2}^\infty {1 \over (n+1)!} \omega ( \Phi, L_{n}(\underbrace{\Phi, \cdots, \Phi}_{n \text{ times}})).\label{eq:action}
\end{align}
The notation follows what can be found in string field theory literature, for example \cite{Erler}. Here $\Phi$ is the dynamical field of the theory. It is an element of a graded vector space $\mathcal{H}$. There are at least two gradings on $\mathcal{H}$, an integer cohomological grading and an even/odd grading describing whether elements are commuting or anticommuting. We take the dynamical field $\Phi$ to be grade zero and commuting. In the context of BV quantization $\Phi$ is commuting but extended to include components at all integer cohomological grades. The other components represent the full spectrum of fields and antifields. With other ingredients further described below, the action will then satisfy the classical BV master equation \cite{Zwiebach}. 

The BV inner product $\omega: \mathcal{H}^{\otimes 2} \to \mathcal{H}^0$ is a nondegenerate bilinear form with the exchange property
\begin{align}
	\omega(A, B) = - (-1)^{|A||B|} \omega(B,A).
\end{align}
%for $A,B \in \mathcal{H}$. 
Here and throughout $|A|$ is even or odd if $A\in\H$ is commuting or anticommuting. The BV inner product carries grade $-1$ and is assumed to be anticommuting. The exchange property implies that $\omega$ is a symplectic inner product. In fact, this is the odd symplectic structure which defines the BV antibracket. However we will refer to $\omega$ as the BV inner product to avoid confusion with the phase space symplectic structure to be discussed below. 

The action depends on a sequence of multilinear maps  $L_1=Q, L_2, L_3, \cdots$,
\begin{align}
	L_n : \mathcal{H}^{\otimes n} \to \mathcal{H},
\end{align}
which carry grade $+1$ and anticommute. The maps $L_n$ can be thought of as defining $n$-fold multiplication of vectors in~$\H$ and are referred to as {\it products}. All products are graded symmetric, 
\begin{align}
L_n(\cdots ,A,B,\cdots) = (-1)^{|A||B|}L_n(\cdots ,B,A,\cdots), \label{eq:symmetric}
\end{align}
%for $A,B\in \H$. 
and satisfy a hierarchy of quadratic identities known as {\it $L_\infty$ relations}, the first of which says that the operator $Q$ (a $1$-fold product) is nilpotent, $Q^2=0$. The products also satisfy a relation analogous to ``integration by parts'' known as {\it cyclicity},
\begin{align}
	\omega(A_1, L_{n}(A_2, \cdots A_{n+1}))= -(-1)^{|A_1|} \omega(L_n( A_1,\cdots, A_n) ,A_{n+1}).
\end{align}
%where $A_1,...,A_{n+1}\in\H$. 
A hidden assumption behind cyclicity is that the integral of a total derivative in spacetime does not generate boundary contributions. Often this is taken for granted but we will need to be careful about this. If all conditions outlined above are satisfied, the triplet $(\H,\omega,L_n)$ defines a {\it cyclic $L_\infty$ algebra}. 

\begin{figure}[t]
	\centering
	\includegraphics[scale=.85]{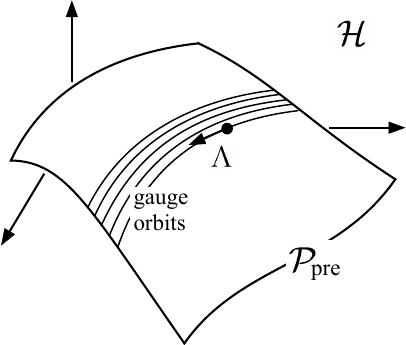}
\caption{Pre-phase space is a nonlinear subspace of $\H$ foliated by a set of gauge orbits whose tangent vectors satisfy \eq{LiePhi}.}\label{fig:CovL3}
\end{figure} 

Varying the action leads to equations of motion
\begin{align}
	Q \Phi + \sum_{n=2}^\infty {1 \over n!} 
	L_n(\underbrace{\Phi, \cdots, \Phi}_{n \text{ times}}) = 0. \label{eq:EOM}
\end{align}
Solutions to the equations of motion define the~\emph{pre-phase space}, denoted by $\mathcal{P}_{\text{pre}}$. Since $\Phi \in \mathcal{H}$, the pre-phase space is a (nonlinear) subspace of $\mathcal{H}$. See figure \ref{fig:CovL3}. Many solutions in pre-phase space are related by gauge transformation and are physically equivalent. The infinitesimal gauge transformation takes the form
\begin{align}
	\Phi\,\to \, \Phi + Q_\Phi \Lambda, \label{eq:gauge_trans}
\end{align}
where $\Lambda$ is an anticommuting element of $\mathcal{H}$ of grade $-1$ and
\begin{align} 
	Q_\Phi A \equiv QA + \sum_{n=1}^\infty {1 \over n!} L_{n+1}(\underbrace{\Phi, \cdots, \Phi}_{n \text{ times}}, A),
\end{align}
%for $A\in \H$ 
is the kinetic operator of the action expanded around the solution $\Phi$. This operator is cyclic, 
\begin{align}
\omega(Q_\Phi A,B) + (-1)^{|A|}\omega(A,Q_\Phi B) = 0,
\end{align}
if boundary contributions can be ignored. Furthermore, if $\Phi$ is a solution, the kinetic operator is nilpotent,
\begin{align}
Q_\Phi^2=0.
\end{align}
These properties hold on account of $L_\infty$ relations and cyclicity of the products. The infinitesimal gauge transformation by $\Lambda$ generates a flow in the pre-phase space represented by a vector field which we also denote as $\Lambda$. The Lie derivative along this vector field satisfies
\begin{align}
	\mathcal{L}_\Lambda \Phi = Q_\Phi \Lambda. \label{eq:LiePhi}
\end{align}
If physically equivalent solutions are counted only once, the pre-phase space $\mathcal{P}_{\text{pre}}$ is projected down to {\it phase space}, which we denote as $\mathcal{P}$. Phase space is the space of inequivalent physical states of the theory, or, in the covariant phase space formalism, it is the space of inequivalent classical solutions. 

To discuss the symplectic structure on phase space we need differential forms. These can be introduced via the exterior derivative on pre-phase space, denoted by $\delta$. This is nilpotent, 
\begin{align}
\delta^2=0,
\end{align}
and generates a complex of differential forms on $\Ppre$. After the suitable quotient, it further generates a complex of differential forms on phase space $\P$. We assume that $\delta$ anticommutes through anticommuting objects in $\H$ and carries grade zero. The exterior derivative of a solution creates a differential $\delta\Phi$ which is anticommuting and carries grade $0$. Note that $\delta\Phi$ depends on the solution $\Phi$. This is because the variations of a solution which preserve the equations of motion depend on the solution itself. Applying the exterior derivative to the equations of motion implies that $\delta\Phi$ satisfies the constraint
\begin{align}
Q_\Phi \big(\delta\Phi\big) =0,  \label{eq:QdeltaPhi}
\end{align}
which depends explicitly on $\Phi$. 

The Lie derivative along a vector field $v$ on pre-phase space is given according to Cartan's magic formula
\begin{equation}
	\mathcal{L}_v = \iota_v\delta+\delta\iota_v,\label{eq:Cartan}
\end{equation}
where $\iota_v$ denotes contraction with $v$. We take the convention that contraction places the vector field in the first entry of the differential form. Contraction is an anticommuting operation at grade zero. From \eq{QdeltaPhi}  we learn that contracting $\delta\Phi$ with a vector field creates a solution to the linearized equations of motion in the background $\Phi$. Specifically contracting with the the vector field of a gauge transformation creates a linearized pure gauge solution 
\begin{equation}
	\iota_\Lambda\delta\Phi = Q_\Phi\Lambda,
\end{equation}
as follows from \eq{LiePhi}. The Lie derivative allows us to define gauge transformations of the entire complex of differential forms on $\Ppre$. For example, the infinitesimal gauge transformation of $\delta\Phi$ is given by 
\begin{align}
\mathcal{L}_\Lambda\delta\Phi = \delta(\mathcal{L}_\Lambda\Phi) = \delta(Q_\Phi \Lambda) = L_2^{\Phi}(\Lambda,\delta\Phi),
\end{align}
where 
\begin{align}
L_2^\Phi(A,B) \equiv L_2(A,B) + \sum_{n=1}^\infty \frac{1}{n!}L_{n+2}(\underbrace{\Phi,\cdots,\Phi}_{n\text{ times}},A,B) \, ,
\end{align}
is the $2$-fold product of fluctuations around the solution $\Phi$. If boundary terms can be ignored, the product is cyclic:
\begin{align}
\omega\big(L_2^\Phi(A,B),C\big) + (-1)^{|A|}\omega\big(A,L_2^\Phi(B,C)\big) = 0.
\end{align}
We emphasize that $\Phi$ and the differential $\delta \Phi$ must be modified simultaneously by the gauge transformation, otherwise \eq{QdeltaPhi} is violated. 

\subsection{Symplectic form}
\label{subsec:Omega}

The existence of a symplectic structure on phase space $\P$ requires that $\Phi$ depends on at least one real parameter $t$ which can be interpreted as time. The field could additionally depend on any number of spatial coordinates. However, we assume that spatial dimensions are compact. This avoids questions about boundary conditions and boundary terms. We touch on these issues briefly in the context of general relativity in subsection~\ref{subsec:GR}.

\begin{figure}[t]
	\centering
	\includegraphics[scale=1.75]{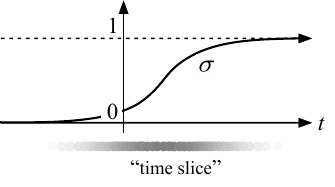}
	\caption{\label{fig:CovL1} The change of the sigmoid between 0 and 1 creates a ``fuzzy" version of a time slice.}
\end{figure}

The proposed symplectic structure is
\begin{align}
\Omega = \frac{1}{2}\omega\big(\delta\Phi,[Q_\Phi,\sigma]\delta\Phi\big).\label{eq:Omega2}
\end{align}
As presented, $\Omega$ is a 2-form on pre-phase space $\Ppre$. Sometimes this is called the {\it pre-symplectic structure}. But gauge invariance, further explained below, guarantees that $\Omega$ is also a symplectic form on phase space~$\P$. The symplectic structure depends on a commuting, grade 0 operator
\begin{align}
\sigma:\H\to \H,
\end{align}
called the {\it sigmoid}. The sigmoid has two properties. First, it is preserved through the BV inner product, 
\begin{align}
	\omega(\sigma A, B)  =\omega(A, \sigma B),\label{eq:sigmalocal}
\end{align}
which implies
\begin{align}
	\omega\big([Q_\Phi,\sigma] A, B\big)  = (-1)^{|A|}\omega\big(A, [Q_\Phi,\sigma] B\big) \, .
\end{align}
This property is forced because any part of $\sigma$ which is not preserved would not contribute to the symplectic form. The second property is a pair of boundary conditions in the infinite past and infinite future:  
\begin{align} 
	\lim_{t \to -\infty} \sigma = 0, \quad \quad
	\lim_{t \to \infty} \sigma = 1.\label{eq:sigmalimit}
\end{align}
The sigmoid defines something analogous to a time slice for the symplectic form. The ``time slice" however is not necessarily a Cauchy surface, but something more diffuse and abstract created in the process of the sigmoid's transition from $0$ to $1$ in time. See figure~\ref{fig:CovL1}. The time slice can be made sharp by choosing a sigmoid to act exclusively through multiplication by a unit step function which is zero in the past, and one in the future of a chosen Cauchy surface. If the Lagrangian is local, $\Omega$ will simplify to an integral of the symplectic charge density over the Cauchy surface, reproducing standard results. If the theory is not local the unit step function will not necessarily localize to a Cauchy surface. In this context it makes sense to allow $\sigma$ to be a more general kind of operator. The term ``sigmoid" is inspired by that of a monotonically increasing sigmoid curve as shown in figure \ref{fig:CovL1}, but the sigmoid is not required to increase monotonically and is not even required to act through multiplication by a function. In this context the limits \eq{sigmalimit} do not refer to the dependence of $\sigma$ on time, but to the region of support of the fields on which $\sigma$ acts. 

While $\Omega$ is not necessarily localized to a Cauchy surface it is localized to regions where the sigmoid is changing. This can be understood from the fact that the dependence of $\sigma$ appears in the form of a commutator $[Q_\Phi, \sigma]$. Since both $0$ and $1$ commute with $Q_\Phi$, the symplectic form as expressed by \eq{Omega2} cannot have any contribution from the infinite past and future. This expectation however assumes that the theory is at least somewhat localized in time. Otherwise a change in $\sigma$ at $t=0$ might effect the commutator $[Q_\Phi, \sigma]$ at all times. We will not try to specify what it means to have a sufficiently localized Lagrangian. For a nonlocal theory it may be necessary to check that the symplectic form is finite and possesses all necessary properties. 

\subsection{Consistency}
\label{subsec:consistency}

The symplectic form must satisfy a set of consistency conditions: 
\begin{description}
\item{\bf Closedness:} The symplectic form is closed, 
\begin{equation}
	\delta\Omega = 0.
\end{equation}
\item{\bf Zero tangents vanish:} Any tangent vector $\Lambda$ to pre-phase space which generates a gauge transformation acts as zero on the symplectic form,
\begin{equation}
	\iota_\Lambda\Omega = 0.
\end{equation}
This condition is necessary because the symplectic structure is supposed to be well-defined on phase space. Since gauge equivalent solutions are identical in phase space, the vector fields which generate gauge transformations must vanish. 
\item{\bf Gauge invariance:} The symplectic form is gauge invariant,
\begin{align}
\L_\Lambda\Omega = 0.
\end{align}
Again this is necessary because the symplectic structure must be well-defined on phase space. This condition follows immediately from the previous two and Cartan's magic formula~\eq{Cartan}. 
\item{\bf Conservation:} The symplectic form is independent of changes in $\sigma$ which preserve the boundary conditions \eq{sigmalimit}. This is an extension of the standard statement that the symplectic structure is independent of the choice of Cauchy surface in the covariant phase space formalism.
\item{\bf Nondegeneracy:} Any tangent vector $v$ to pre-phase space which annihilates the symplectic form,
\begin{align}
\iota_v\Omega=0,
\end{align}
must generate a gauge transformation. Therefore all tangents in the kernel of $\Omega$ are equivalent to zero, and the symplectic form on phase space is nondegenerate.
\end{description}
We will show that all these conditions are satisfied except for the last. Nondegeneracy is sometimes taken to hold by declaring that any zero mode of $\Omega$ is a gauge transformation by definition.  From the point of view of our development this does not seem sufficient because gauge transformations are already defined by~\eq{LiePhi}. A better argument is to show that the symplectic form can be inverted through the Poisson (or Peierls~\cite{Peierls}) bracket. This however is outside the scope of what we discuss in this article.

Before moving forward we must address an important problem. The symplectic structure is naively zero. This is clear once we remember that $Q_\Phi$ is cyclic and annihilates~$\delta\Phi$. Explicitly, 
\begin{align}\Omega & = \frac{1}{2}\Big[\omega\big(\delta\Phi,Q_\Phi\sigma \delta\Phi\big)- \omega\big(\delta\Phi,\sigma Q_\Phi\delta\Phi\big)\Big]\nonumber\\
& = \frac{1}{2}\Big[\omega\big(Q_\Phi \delta\Phi,\sigma \delta\Phi\big)- \omega\big(\delta\Phi,\sigma Q_\Phi\delta\Phi\big)\Big]\nonumber\\
& = 0.\ \ \label{eq:vanish}
\end{align}
The problem with this argument is that the two terms of the commutator are not separately localized to finite time. So in a sense each term is ``zero" multiplied by a divergent factor caused by integration over the infinite volume of spacetime. Another point of view is that cyclicity of $Q_\Phi$ breaks down. Total derivatives create boundary contributions which do not vanish because the fields do not vanish at infinity. We can sidestep all of these problems by declaring that the operator $[Q_\Phi,\sigma]$ must be treated as a single unit which cannot be decomposed inside the BV inner product. Nevertheless it is worthwhile to look at \eq{vanish} more carefully and determine the missing boundary terms. This seems to require knowing the derivative structure of $Q_\Phi$, but there is a work-around using what will be called {\it tau regularization}. The idea is to introduce a commuting, grade zero operator,
\begin{align}
\tau: \H\to \H,
\end{align}
similar to the sigmoid but satisfying conditions
\begin{align}
\tau = 1\ \ (\mathrm{finite}\ t),\ \ \ \ \ \ \lim_{t\to\pm\infty}\tau = 0.\label{eq:tau_bdry}
\end{align}
This is a slight abuse of notation since technically these conditions are incompatible. What should be understood by $\tau$ is a limit of a sequence of operators $\tau_n$ satisfying 
\begin{align}
\lim_{n\to\infty}\tau_n = 1\ \ (\mathrm{finite}\ t),\ \ \ \ \ \ \lim_{t\to\pm\infty}\tau_n = 0.
\end{align}

\begin{figure}[t]
	\centering
	\includegraphics[scale=1]{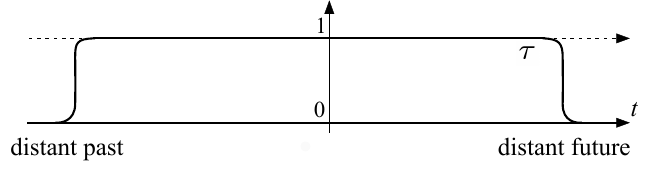}
	\caption{The tau regulator is implemented by an operator $\tau$ which is effectively $1$ at finite time but vanishes in the distant past and future.}\label{fig:CovL2}
\end{figure} 

\noindent The operator $\tau$ can be imagined as a table-shaped curve (hence ``tau" for table) which rises from 0 to 1 in the distant past and falls back from 1 to 0 in the distant future. See figure \ref{fig:CovL2}. Since $\tau$ acts as the identity operator at finite time, we can insert it in front of $\Phi$ and/or $\delta\Phi$ in the symplectic form without changing the result. However when $\tau$ is present we can break up the commutator $[Q_\Phi,\sigma]$ because volume divergences are consistently regularized. We can also apply cyclicity without generating boundary terms because the fields vanish at infinity. For a generic nonlocal theory each factor of $\delta\Phi$ and $\Phi$ in the symplectic form will come with its own spacetime integral, and inserting $\tau$ in each factor will ensure that every integrand vanishes at infinity.  However this is probably more than needed. If the Lagrangian possesses a reasonable degree of locality, vanishing at infinity in one integrand will be sufficient to ensure vanishing at infinity in all. For the sake of present discussion it is convenient to insert one $\tau$ in front of the first $\delta\Phi$ in the symplectic form, giving 
\begin{align}
\Omega = \frac{1}{2}\omega\big(\tau\delta\Phi,[Q_\Phi,\sigma]\delta\Phi\big).\label{eq:Omega_reg}
\end{align}
Now we are free to use cyclicity. Repeating the calculation of \eq{vanish} we find, instead of zero, 
\begin{equation}
\Omega = \frac{1}{2}\omega\big([Q_\Phi,\tau] \delta\Phi,\sigma\delta\Phi\big).\label{eq:Omega_bdry}
\end{equation}
By the same reasoning as with the sigmoid, the commutator $[Q_\Phi,\tau]$ is localized to where $\tau$ is changing. By \eq{tau_bdry} this is the infinite past and future, so in fact this is precisely the boundary term which is missing from \eq{vanish}. To understand its interpretation we write $\tau$ as
\begin{align}
\tau = \sigma_--\sigma_+,
\end{align}
where $\sigma_-$ is a sigmoid operator which shifts from $0$ to $1$ in the distant past and $\sigma_+$ is a sigmoid operator which shifts from $0$ to $1$ in the distant future. Like $\tau$, these sigmoids should be understood as a limit. Plugging this in gives
\begin{align}
\Omega &= -\frac{1}{2}\omega\big(\delta\Phi,[Q_\Phi,\tau] \sigma\delta\Phi\big)\nonumber\\
& = -\frac{1}{2}\omega\Big(\delta\Phi,\big[Q_\Phi,\sigma_--\sigma_+\big] \sigma\delta\Phi\Big)\nonumber\\
& = -\frac{1}{2}\omega\Big(\delta\Phi,\big[Q_\Phi,\sigma_-\big]\sigma\delta\Phi\Big)+\frac{1}{2}\omega\big(\delta\Phi,[Q_\Phi,\sigma_+] \sigma\delta\Phi\big).
\end{align}
The commutator $[Q_\Phi,\sigma_-]$ is localized in the infinite past. Here the original sigmoid $\sigma$ vanishes, so the first term is zero. The commutator $[Q_\Phi,\sigma_+]$ is localized in the infinite future. Here $\sigma$ is equal to 1. Therefore we find that
\begin{align}
\Omega = \frac{1}{2}\omega\big(\delta\Phi,[Q_\Phi,\sigma_+]\delta\Phi\big).
\end{align}
This is the symplectic form itself with a specific choice of sigmoid which localizes to the distant future. In this way the symplectic form is purely a boundary term. But this is a trivial expression of the fact that the time slice can be pushed as far as we want into the future. 

We are ready to prove the consistency of the symplectic form. We start with closedness. This will require opening up the commutator $[Q_\Phi,\sigma]$ and invoking cyclicity, so the tau regulator is needed. In the present circumstance it is most convenient to insert $\tau$ in front of all instances of $\Phi$ and $\delta\Phi$, so the symplectic form is expressed as 
\begin{align}
\Omega = \frac{1}{2}\omega\big(\tau\delta\Phi,[Q_{\tau\Phi},\sigma]\tau\delta\Phi\big).
\end{align}
Applying the exterior derivative $\delta$ turns the commutator with $Q_{\tau\Phi}$ into a commutator with $L_2^{\tau\Phi}$:
\begin{align}
\delta\Omega = -\frac{1}{2}\omega\Big(\tau\delta\Phi,L_2^{\tau\Phi}(\tau\delta\Phi,\sigma\tau\delta\Phi\big)\Big)+\frac{1}{2}\omega\Big(\tau\delta\Phi,\sigma L_2^{\tau\Phi}(\tau\delta\Phi,\tau\delta\Phi\big)\Big).
\end{align}
The second term vanishes because $\delta\Phi$ is anticommuting and the product is symmetric by \eq{symmetric}. The tau regulator allows us to use cyclicity on the first term to find
\begin{align}
\delta\Omega = -\frac{1}{2}\omega\Big(L_2^{\tau\Phi}(\tau\delta\Phi,\tau\delta\Phi\big),\sigma\tau\delta\Phi\Big).
\end{align}
This again is zero by symmetry of the product, which completes the proof of closedness. 

Next we show that zero tangents vanish. This does not require the tau regulator. Contracting with the vector field of a gauge transformation gives 
\begin{equation}
\iota_\Lambda \Omega = -\omega\big(Q_\Phi \Lambda,[Q_\Phi,\sigma]\delta\Phi\big).
\end{equation}
Using cyclicity we can move $Q_\Phi$ off of $\Lambda$ to act on the second entry of the BV inner product. We do not encounter boundary terms in the process because $[Q_\Phi,\sigma]$ vanishes at infinity. When $Q_\Phi$ acts on the second entry of the BV inner product it gives zero because $Q_\Phi$ is nilpotent and annihilates~$\delta\Phi$. This establishes that zero tangents vanish. Gauge invariance then follows from Cartan's magic formula. 

Finally we check conservation. Suppose we have two symplectic forms defined with sigmoids $\sigma$ and $\sigma'$. Their difference is
\begin{align}
\Omega(\sigma)- \Omega(\sigma') = \frac{1}{2}\omega\Big(\delta\Phi,\big[Q_\Phi,(\sigma-\sigma')\big]\delta\Phi\Big).\label{eq:Omegasigmasigmap}
\end{align}
The difference between two sigmoids vanishes in the infinite past and future. Therefore there is no volume divergence, and we can use cyclicity without generating boundary terms to arrange $Q_\Phi$ to operate on $\delta\Phi$. Then we find zero. 

\section{Examples} 
\label{sec:examples}

In this section we compute the symplectic form in examples. The first issue that arises is that most ordinary field theories are not expressed in the language of $L_\infty$ algebras. While it is always possible to construct such a description, this may require some effort. Fortunately, if our goal is only to {\it evaluate} the symplectic form, most of this effort is not necessary. This is because the symplectic form only depends on $Q_\Phi$ when it acts on fields at grade~0. This information is easily extracted from the free action for fluctuations around the solution~$\Phi$. Finding out how the $L_\infty$ products act on fields at other grades requires more work. This is relevant for understanding how the theory's gauge structure is formalized by a cyclic $L_\infty$ algebra and for solving the BV master equation. But, as will be clear in examples, the symplectic structure is automatically gauge invariant. We do not need the gauge transformations expressed in $L_\infty$ form to see this.  

Still a little more is needed to convert the formula \eq{Omega} into something which is ready to evaluate for a typical field theory. For this we introduce a basis $e_i$ for the grade 0 subspace of $\H$ and a dual basis $e^i$ for the grade 1 subspace so that 
\begin{align}
A\in \H,\ \  \mathrm{grade}\ 0 \ \ \ \longrightarrow\ \ \ A &= A^i e_i, \nonumber\\
B\in \H,\ \  \mathrm{grade}\ 1\ \ \ \longrightarrow\ \ \ B & = B_i e^i,
\end{align}
where $A^i$ and $B_i$ are coefficient fields. We use DeWitt notation where the index $i$ includes spacetime coordinate as well as internal indices. The basis vectors $e_i$ at grade zero are commuting and the dual basis vectors $e^i$ at grade 1 are anticommuting. The bases are dual in the sense that their BV inner product satisfies:
\begin{align}
\omega(e_i,e^j) = \delta_i^j \, .
\end{align}
We further define
\begin{align}
Q_\Phi e_i = Q_{ij}e^j,\ \ \ \ \sigma e_i = \sigma_i^j e_j, \ \ \ \ \sigma e^i  = \sigma^i_j e^j.\label{eq:sigmae^i}
\end{align}
Cyclicity of $Q_\Phi$ implies that the matrix elements $Q_{ij}$ are symmetric,
\begin{align}
Q_{ij}=Q_{ji},
\end{align}
while the condition that the sigmoid is preserved through the BV inner product \eq{sigmalocal} implies that the matrix elements $\sigma_i^j$ in the last two equations of \eq{sigmae^i} are the same. Consider the free action for the fluctuation $\varphi$ of a solution $\Phi$. If expanded in a basis $\varphi = \varphi^ie_i$, the definitions above imply
\begin{align}
S_\mathrm{fluctuation} = -\frac{1}{2}\omega(\varphi,Q_\Phi\varphi) = -\frac{1}{2}\varphi^i Q_{ij}\varphi^j.\label{eq:Sfluctuation}
\end{align}
We can also write  $\Phi = \Phi^i e_i$ and expand the symplectic form in a basis 
\begin{align}
\Omega = \frac{1}{2}\omega\big(\delta\Phi,[Q_\Phi,\sigma]\delta\Phi\big)= -\frac{1}{2}\delta\Phi^i \big(Q_{ik}\sigma^k_j - \sigma_i^k Q_{kj}\big)\delta\Phi^j.\label{eq:Omegabasis}
\end{align}
Note the sign on the final equation. The purpose of writing everything in components is the following.  Even if our action is not expressed in $L_\infty$ form we can always expand it around a solution. Possibly after integrating by parts, the free action for the fluctuation field can always be expressed as the final equation of \eq{Sfluctuation}. From this we learn the matrix elements $Q_{ij}$. Then, once we have made a choice of sigmoid, we can compute the symplectic form from the final equation of \eq{Omegabasis}. This will be our alternative to the traditional covariant phase space method for computing the symplectic structure. 

\subsection{Scalar field theory}
\label{subsec:scalar}

Our first example is a real scalar field $\phi$ with a potential $V(\phi)$
\begin{align}
S = \int d^Dx \left(-\frac{1}{2}\d_\mu\phi \d^\mu\phi -V(\phi)\right).\label{eq:action_scalar}
\end{align}
We work in flat space without spatial boundary. We can expand the action in terms of a fluctuation $\varphi$ around a classical solution $\phi$. The free action for the fluctuation field is 
\begin{align}
S_\mathrm{fluctuation} = -\frac{1}{2}\int d^Dx\, \varphi\Big(-\Box +V''(\phi)\Big)\varphi,
\end{align}
where $\Box= \d_\mu\d^\mu$ is the d'Alembert operator. From this we can read the matrix elements $Q_{ij}$:
\begin{equation}
Q(x,y) = \Big(-\Box +V''(\phi)\Big)\delta^D(x-y).
\end{equation}
We choose a sigmoid which acts through multiplication by a function $\sigma(x)$. The matrix $\sigma_i^j$ would then be
\begin{align}
\sigma(x,y) = \sigma(x)\delta^D(x-y).
\end{align}
After integrating the delta functions it is easy to see that the commutator of $Q_{ij}$ and $\sigma_i^j$ essentially amounts to the commutator of the Klein-Gordon operator with the sigmoid. This is straightforward to compute
\begin{align}
\big[-\Box+V''(\phi),\sigma] = -2(\d_\mu\sigma)\d^\mu -(\Box\sigma).\label{eq:Qsigma_scalar}
\end{align}
Substituting into \eq{Omegabasis} gives the symplectic form
\begin{align}
\Omega = -\frac{1}{2}\int d^D x\, \delta \phi\Big(-2(\d_\mu\sigma)\d^\mu -(\Box\sigma)\Big)\delta\phi.
\end{align}
Because $\delta\phi$ squares to zero we can drop the $\Box \sigma$ term to write
\begin{align}
\Omega = \int d^D x\, \big(\d_\mu\sigma\big)\,\delta\phi\,\d^\mu\delta \phi.\label{eq:phiOmega}
\end{align}
The sigmoid comes with a derivative so the integrand is localized to where the sigmoid is changing. This is how the sigmoid creates a ``diffuse'' time slice. The time slice can be made sharp by assuming that the sigmoid is a step function which jumps from $0$ to $1$ at $t=0$. The derivative of the sigmoid is then a delta function which eliminates the integration over time,
\begin{equation}
	\d_\mu \sigma(x) = \d_\mu \theta(t) = \delta(t)\delta_\mu^t. 
\end{equation}
The symplectic form becomes
\begin{align}
\Omega = - \int_{t=0} d^{D-1} x \, \delta\phi\, \frac{\d(\delta\phi)}{\d t},
\end{align}
where the sign appears from the mostly plus metric convention. The canonical momentum conjugate to the field $\phi$ is
\begin{align}
\pi(x) = \frac{\d \mathcal{L}(x)}{\d\left(\frac{\d\phi(x)}{\d t}\right)} = \frac{\d \phi(x)}{\d t},
\end{align}
where $\mathcal{L}(x)$ is the Lagrangian density in \eq{action_scalar}. The symplectic form is finally expressed as
\begin{align}
\Omega = \int_{t=0} d^{D-1}x\, \delta\pi\, \delta\phi,
\end{align}
which has the expected structure $\delta p_i\wedge\delta q^i$.

The factor which multiplies the sigmoid in \eq{phiOmega} may be recognized as the so-called symplectic current density,
\begin{align}
\omega^\mu = \delta\phi\, \d^\mu\delta \phi.
\end{align}
Because the exterior derivative on pre-phase space preserves the equations of motion, $\delta\phi$ must satisfy~\eq{QdeltaPhi}. For scalar field theory this reduces to
\begin{align}
\big(-\Box + V''(\phi)\big)\delta\phi = 0.\label{eq:Qdeltaphi}
\end{align}
From this it follows that the symplectic current density is conserved:
\begin{align}
\d_\mu \omega^\mu & = (\d_\mu\delta\phi)( \d^\mu\delta\phi) + \delta\phi \Box\delta\phi\nonumber\\
& = \delta\phi \Big(V''(\phi)\delta\phi\Big) \nonumber\\
& = 0 \, .
\end{align}
In the second step we used  \eq{Qdeltaphi} and that $\d_\mu\delta\phi$ squares to zero. In the last step we used that $\delta\phi$ squares to zero. This explains why the symplectic structure is independent of the sigmoid. Suppose $\sigma$ and $\sigma'$ are two sigmoids satisfying the boundary conditions~\eq{sigmalimit}. The difference of the respective symplectic forms is
\begin{align}
\Omega(\sigma) - \Omega(\sigma') = \int d^D x\, \d_\mu(\sigma-\sigma') \omega^\mu.
\end{align}
Because the difference of sigmoids vanishes at infinity we can integrate by parts without generating boundary terms. Then the fact that the symplectic current is conserved implies that the symplectic structures are equal. 

Let us understand why the symplectic form is not zero. If the commutator is not evaluated first as \eq{Qsigma_scalar}, the symplectic form appears as 
\begin{align}
\Omega = -\frac{1}{2}\int d^Dx\, \delta\phi\big[-\Box+V''(\phi),\sigma\big]\delta\phi.
\end{align}
The second term of the commutator can be said to vanish on account of \eq{Qdeltaphi}. After integration by parts twice,  the first term naively vanishes for the same reason. However, integration by parts generates boundary terms. Stoke's theorem gives the boundary terms as 
\begin{align}
\Omega  = \bigg( \int_{t=\infty} & +\int_{t=-\infty}\bigg)d^{D-1}x\, n_\mu \Big(\sigma \delta\phi\, \d^\mu\delta\phi\Big) ,
\end{align}
where $n_\mu$ is the outward pointing unit normal on the past and future boundary (spatial boundaries are not present). The boundary term from $t=-\infty$ does not contribute because the sigmoid vanishes. On the boundary at $t=+\infty$ the outward pointing normal points to the future
\begin{align}
n_\mu = \delta_\mu^t.
\end{align}
Because the sigmoid is equal to one we obtain 
\begin{align}
\Omega = - \int_{t=\infty} d^{D-1} x \, \delta\phi\, \frac{\d(\delta\phi)}{\d t}.
\end{align}
This is the symplectic form itself with the time slice pushed to infinity. The conclusion is the same as what we argued in subsection \ref{subsec:consistency}.

\subsection{Yang-Mills theory}
\label{subsec:YM}

The next example is Yang-Mills theory. This is a more serious test of the formula because Yang-Mills theory has derivative interactions. Derivative interactions are needed to distinguish the commutator with $Q_\Phi$ from the commutator with $Q$. This is what gives the symplectic form an interesting nonlinear structure, and is one important respect that our formula goes beyond Witten's proposal \cite{Witten}.

The Yang-Mills action is 
\begin{align} 
S = - {1 \over 4 g^2} \int d^D x \, \Tr\big[ F_{\mu\nu} F^{\mu\nu} \big] .\label{eq:YMaction}
\end{align}
Again we work in flat space without spatial boundary. The Yang-Mills coupling constant is $g$ and the field strength $F_{\mu \nu}$ is derived from the gauge field $A_\mu$ as
\begin{align}
	F_{\mu \nu} = \d_\mu A_\nu- \d_\nu A_\mu - i [A_\mu, A_\nu].
\end{align}
The gauge field $A_\mu$ takes values in a Lie algebra. The commutator and trace refer to the Lie bracket and Killing form in this algebra. Sometimes it is clearer to express the gauge field in terms of Lie algebra generators $T^a$ as
\begin{align}
A_\mu = A_\mu^a T^a,
\end{align}
where $A_\mu^a$ are number-valued coefficients and $a$ is a Lie algebra index. We normalize $\Tr[T^aT^b] = \delta^{ab}$.

The free action for a fluctuation $a_\mu$ around a solution $A_\mu$ is 
\begin{align} \label{eq:3.17}
	S_\mathrm{fluctuation} =  - {1 \over 2 g^2}\int d^D x\,
	\Tr \big[ a_\mu \Box^{\mu\nu}a_\nu \big],
\end{align}
where 
\begin{align}
\Box^{\mu\nu} = \mathcal{D}^{\nu} \mathcal{D}^{\mu} - \eta^{\mu \nu} \mathcal{D}^2 \, ,
\end{align}
and 
\begin{align}
	\mathcal{D}_\mu B \equiv \d_\mu B - i [A_\mu,B] \, ,
\end{align}
is the gauge covariant derivative in the adjoint representation. To extract the matrix elements $Q_{ij}$ it is helpful to employ Lie algebra indices. Let $\mathcal{O}$ be an operator acting on Lie-algebra valued field $B = B^a T^a$. We can define matrix elements $\mathcal{O}^{ab}$ according to 
\begin{align}
\mathcal{O}B = (\mathcal{O}^{ab}B^b)T^a.
\end{align}
Then $Q_{ij}$ may be written as
\begin{align}
Q^{\mu\nu,ab}(x,y) = \frac{1}{g^2}\Box^{\mu\nu,ab}\delta^D(x-y).
\end{align}
If the sigmoid acts through multiplication by $\sigma(x)$, the $\sigma^i_j$ may be written as
\begin{align}
\sigma^{\mu\ ,ab}_{\ \nu}(x,y) = \sigma(x)\delta^\mu_{\ \nu} \delta^{ab}\delta^D(x-y).
\end{align}
After integrating the delta functions and summing the Lorentz and Lie algebra indices, the commutator of $Q_{ij}$ with $\sigma^i_j$ reduces to the commutator 
\begin{align}
\big[\Box^{\mu\nu},\sigma]=(\d^\mu\d^\nu\sigma)-\eta^{\mu\nu}(\Box\sigma) +(\d^\mu\sigma) \mathcal{D}^\nu+(\d^\nu\sigma) \mathcal{D}^\mu - 2\eta^{\mu\nu} (\d_\lambda \sigma) \mathcal{D}^\lambda.
\end{align}
This is the form of the operator $[Q_\Phi,\sigma]$ in Yang-Mills theory. It depends on $\Phi$ through the appearance of $A_\mu$ in the gauge covariant derivative. Substituting this into \eq{Omegabasis}, contributions with two derivatives of the sigmoid drop out for symmetry reasons. Factoring out $\d_\mu\sigma$ from the remaining terms gives 
\begin{align}
\Omega = \int d^D x \big(\d_\mu\sigma\big) \Pi^\mu, \label{eq:YMPiOmega}
\end{align}
where
\begin{align}
\Pi^\mu = -\frac{1}{2 g^2}\Tr\Big[\delta A^\mu\mathcal{D}^\nu \delta A_\nu + \delta A^\nu \mathcal{D}_\nu \delta A^\mu -2 \delta A^\nu \mathcal{D}^\mu \delta A_\nu\Big].
\end{align}
If we expect to obtain the correct result, $\Pi^\mu$ should be equivalent to the standard symplectic current density of Yang-Mills \cite{Crnkovic}
\begin{align}
\omega^\mu &= -\frac{1}{g^2}\Tr\big[ \delta F^{\mu\nu}\delta A_\nu\big],\ \ \ \ \ \ \delta F_{\mu\nu}= \mathcal{D}_\mu \delta A_\nu - \mathcal{D}_\nu \delta A_\mu.
\end{align} 
The two are related by an improvement term 
\begin{align}
\Pi^\mu = \omega^\mu +\d_\nu B^{\mu\nu},
\end{align}
given by an antisymmetric tensor
\begin{align}
B^{\mu\nu} = -\frac{1}{2g^2}\Tr\big[\delta A^\mu\delta A^\nu\big].
\end{align}
The improvement term contributes a total derivative to the integrand 
\begin{align}
\d_\nu\big((\d_\mu\sigma) B^{\mu\nu}\big)=(\d_\nu\d_\mu\sigma) B^{\mu\nu} +(\d_\mu\sigma)\d_\nu B^{\mu\nu} = (\d_\mu\sigma)\d_\nu B^{\mu\nu}. \label{eq:Btotalderivative}
\end{align}
which integrates to zero because $\d_\mu\sigma$ vanishes in the infinite past and future (we do not consider spatial boundaries). Therefore we reproduce the correct symplectic structure
\begin{align}
\Omega = -\frac{1}{g^2}\int d^Dx\, \d_\mu\sigma \Tr\big[ \delta F^{\mu\nu}\delta A_\nu\big],
\end{align}
including the needed dependence on $A_\mu$ generated through derivative interactions. One can show without too much difficulty that symplectic structure is gauge invariant and that zero tangents vanish \cite{Crnkovic}. These arguments do not require that Yang-Mills gauge symmetry is formulated in terms of an $L_\infty$ algebra. However, an explicit $L_\infty$ description of Yang-Mills theory has been worked out in \cite{Hohm,Zeitlin}. 

\subsection{General relativity}
\label{subsec:GR}

Now we turn to general relativity. The novelty of general relativity relative to previous examples is that the symplectic structure can receive contribution from the boundary of the Cauchy surface~\cite{Harlow}. However we do not know how the formula \eq{Omega} should be modified in the presence of spatial boundaries. Therefore at first we discuss the symplectic structure without spatial boundary. Then we will see what can be done about boundary effects. 

We consider a spacetime $M$ with metric $g_{\mu\nu}$ and without spatial boundary (initially). The Einstein-Hilbert action is 
\begin{align}
	S= {1 \over 2 \kappa} \int_M \vol  \, R ,
\end{align}
where $R$ is the Ricci scalar, $\kappa = 8\pi G$, and ``$\vol$'' will always denote the canonical volume form of the manifold or hypersurface over which we are integrating (in this case $\vol = \sqrt{-g}\,d^D x$ with $g$ the determinant of the metric in the coordinates $x^\mu$ on $M$). We can expand the action in terms of a fluctuation $h_{\mu\nu}$ around a background solution of Einstein's equation $g_{\mu \nu}$. The free action for the fluctuation turns out to be 
\begin{align}
	S_{\text{fluctuation}} = - {1 \over 8 \kappa} \int_M \vol  \, h_{\mu \nu }\Box^{\mu \nu \alpha \beta} h_{\alpha \beta},
\end{align}
where the kinetic operator is
\begin{align} \label{eq:3.49}
	\Box^{\mu\nu\alpha\beta} = \,
	& 2\nabla^{(\alpha}g^{\beta)(\mu}\nabla^{\nu)}-\big(g^{\mu\nu}\nabla^{(\alpha}\nabla^{\beta)} 
	+ g^{\alpha\beta}\nabla^{(\mu}\nabla^{\nu)}\big)
	\nonumber \\
	& +\big(g^{\mu\nu}g^{\alpha\beta}-g^{\mu(\alpha}g^{\beta)\nu}\big)\nabla^2.
\end{align}
From this we can read off the matrix elements $Q_{ij}$:
\begin{align}
	Q^{\mu \nu \alpha \beta}(x,y) = {\sqrt{-g} \over 4\kappa} \Box^{\mu \nu \alpha \beta} \delta^{D} (x-y).
\end{align}
If the sigmoid acts through multiplication, the matrix $\sigma_i^j$ should take the form
\begin{align}
	\sigma_{\mu\nu}^{\ \ \, \alpha\beta}(x,y) =\sigma(x) \delta^\alpha_{\ (\mu} \delta^\beta_{\ \nu)}  \delta^D(x-y).
\end{align}
After integrating over the delta functions and contracting the spacetime indices, the commutator of $Q_{ij}$ and $\sigma^i_j$ amounts to the commutator of the kinetic operator with the sigmoid. Repeatedly using 
\begin{align}
	[\nabla^\mu\nabla^\nu,\sigma] = (\nabla^\mu\nabla^{\nu} \sigma) + (\nabla^\mu\sigma)\nabla^\nu + (\nabla^\nu\sigma)\nabla^\mu,
\end{align}
the commutator turns out to be
\begin{align}
	[\Box^{\mu\nu\alpha\beta}, \sigma] =
	&2 \Big[ (\nabla^{(\mu}g^{\nu)(\alpha}\nabla^{\beta)} \sigma) 
	+ (\nabla^{(\mu} \sigma ) g^{\nu)(\alpha}\nabla^{\beta)} 
	+ (\nabla^{(\alpha} \sigma) g^{\beta) (\mu } \nabla^{\nu)}
	\Big]
	\nonumber \\
	& - \Big[
	g^{\mu \nu} ( \nabla^{\alpha} \nabla^{\beta} \sigma ) + g^{\alpha \beta} (\nabla^{\mu} \nabla^{\nu} \sigma )
	+ 2 g^{\mu \nu}( \nabla^{(\alpha} \sigma) \nabla^{\beta)}		
	+ 2g^{\alpha \beta}( \nabla^{(\mu} \sigma) \nabla^{\nu)}
	\Big]
	\nonumber \\
	&+ (g^{\mu\nu}g^{\alpha\beta}-g^{\mu(\alpha}g^{\beta)\nu}) \Big[
	( \nabla^2 \sigma) + 2 ( \nabla^\lambda \sigma) \nabla_\lambda 
	\Big].
\end{align}
Substituting into \eq{Omegabasis} we find that terms with two derivatives of $\sigma$ drop out for symmetry reasons. We can factor $\nabla_\mu\sigma$ out of all nonvanishing contributions so that the symplectic structure is expressed as 
\begin{align} 
\Omega =  \int_M \mathrm{vol}\, \big( \nabla_\mu\sigma\big) \Pi^\mu.
\end{align}
We find that
\begin{align}
	\Pi^\mu = -{1 \over 4 \kappa}\bigg[ &-\delta g^{\mu \alpha} \nabla^\beta \delta g_{\alpha \beta} 
	- \delta g_{\alpha\beta} \nabla^\alpha \delta g^{\beta\mu}
	\nonumber \\
	&+ \delta \ln g  \nabla_\alpha \delta g^{\alpha \mu}
	+ \delta g^{\mu \alpha} \nabla_\alpha \delta \ln g
	\nonumber \\
	&+  \delta \ln g \nabla^\mu \delta \ln g
	+ \delta g_{\alpha \beta} \nabla^\mu \delta g^{\alpha \beta} \bigg],\label{eq:GRPi}
\end{align}
where we used
\begin{align}
	\delta g^{\mu \nu} = - g^{\mu \alpha } g^{\nu \beta} \delta g_{\alpha \beta} \, ,
	\quad \quad
	\delta \ln g 
	= g^{\mu \nu} \delta g_{\mu \nu} \, ,
\end{align}
to simplify the expression. 

We can compare this with the known symplectic current density of general relativity \cite{Crnkovic}
\begin{align}
\omega^\mu = -{1 \over 2 \kappa}\bigg[\delta \Gamma^\mu_{\nu \rho} 
	\bigg(\delta g^{\nu \rho}+ {1\over 2} g^{\nu \rho} \delta \ln g\bigg)
	-\delta \Gamma^\nu_{\rho \nu} 
	\bigg(\delta g^{\mu \rho} + {1 \over 2} g^{\mu \rho} \delta \ln g\bigg)\bigg].
\end{align}
After evaluating differentials of the Christoffel symbols,
\begin{align}
	\delta \Gamma^\mu_{\nu \rho} = {1 \over 2} g^{\mu \lambda}
	\bigg[
	\nabla_{\nu} \delta g_{\rho \lambda} 
	+\nabla_{\rho} \delta g_{\nu \lambda}
	- \nabla_\lambda \delta g_{\nu \rho}
	\bigg],
\end{align}
we find that $\Pi^\mu$ differs from the symplectic current density by an improvement term,
\begin{align}
\Pi^\mu = \omega^\mu +\nabla_\nu B^{\mu\nu},
\end{align}
given by an antisymmetric tensor 
\begin{align}
B^{\mu \nu} = -{1 \over 4\kappa} \delta g^{\mu \alpha} g_{\alpha \beta}\, \delta g^{\beta \nu}.
\end{align}
The improvement term contributes a total derivative to the integrand. In the absence of spatial boundaries, the total derivative integrates to zero. Therefore 
\begin{align}
\Omega = \int_M \vol\, \big(\nabla_\mu\sigma\big) \omega^\mu,\label{eq:GROmega}
\end{align}
and we recover the correct symplectic structure of general relativity.

\begin{figure}[t]
	\centering
	\includegraphics[scale=1]{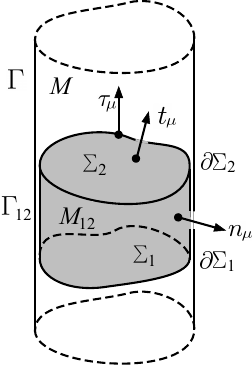}
	\caption{The region $M_{12}$ shaded in gray is where the sigmoid $\sigma=\Theta$ takes the value 1. It is zero everywhere else. $M_{12}$ is bounded in the past and future by Cauchy surfaces $\Sigma_1$ and $\Sigma_2$ and on the spatial boundary by $\Gamma_{12}$. The orientation of $\Gamma_{12}$ and $\Gamma$ is inherited from the boundary of~$M$. The orientation of $\Sigma_1$ and $\Sigma_2$ are determined by the boundary of their past in $M$. Finally, the orientation of $\d \Sigma_1$ and $\d \Sigma_2$ is determined as the boundary of $\Sigma_1$ and $\Sigma_2$. We use $n_\mu$ to denote the outward pointing unit normal on $\Gamma$ and $\Gamma_{12}$. We use $t_\mu$ to denote the future pointing unit normal on both $\Sigma_1$ and $\Sigma_2$. Finally, $\tau_\mu$ denotes the future pointing unit normal on both $\d\Sigma_1$ and $\d\Sigma_2$ which is orthogonal to $n_\mu$, i.e., $n^\mu\tau_\mu = 0$.}\label{fig:CovL5}
\end{figure}

Let us now assume that the spacetime $M$ has a finite spatial boundary~$\Gamma$.\footnote{We assume that general relativity with finite spatial boundaries is meaningful, which is not fully established.} We have no reason to think that our formula works in the presence of spatial boundaries, but nevertheless we can try to adapt what we have learned to this scenario. Our idea is to determine the symplectic structure indirectly through its conservation law. The setup is as follows. We consider a step function sigmoid $\sigma = \Theta$ which operates as an indicator function for the region $M_{12}$ shown in figure \ref{fig:CovL5}. The region is bounded in the past and future by Cauchy surfaces $\Sigma_1$ and $\Sigma_2$, and is contained within the spatial boundary by $\Gamma_{12}$. We require that the sigmoid vanishes everywhere on the boundary of~$M$, so $\Theta$ will have a step discontinuity which approaches the spatial boundary $\Gamma_{12}$ as a limit. Actually, $\Theta$ should not really be called a sigmoid because it does not satisfy the boundary conditions \eq{sigmalimit}. Because it vanishes everywhere on the boundary, there can be no boundary contributions which obstruct cyclicity. Therefore \eq{Omega} will compute {\it zero} instead of the symplectic structure. But it is zero in an interesting form. The step discontinuity on $\Sigma_1$ and $\Sigma_2$ should produce the symplectic charge density integrated over the initial and final Cauchy surfaces. The step discontinuity near $\Gamma_{12}$ should produce the symplectic flux density integrated over $\Gamma_{12}$. With suitable boundary conditions, symplectic flux should be prevented from escaping~$\Gamma_{12}$.  The formula \eq{Omega} will then express conservation of the symplectic structure in the presence of a boundary. From this we can learn the symplectic structure itself.  

Substituting $\sigma=\Theta$ into \eq{GROmega} gives 
\begin{align} \label{3.68}
	0 = \int_{\Sigma_1} \mathrm{vol}\, t_{\mu}\omega^\mu -\int_{\Sigma_2} \mathrm{vol}\, t_{\mu}\omega^\mu -\int_{\Gamma_{12}}\mathrm{vol}\, n_\mu \omega^\mu.
\end{align}
The orientations and the unit normals of these hypersurfaces are defined in figure \ref{fig:CovL5}. The first two terms represent the difference in the symplectic charge between $\Sigma_2$ and $\Sigma_1$, while the last term is the total symplectic flux flowing out of $\Gamma_{12}$. This term should be eliminated with a proper choice of boundary condition. Following \cite{Harlow} we require that the induced metric on $\Gamma$ is fixed, which may be expressed as
\begin{align}
\gamma_{\ \mu}^{\alpha}\gamma_{\ \nu}^{\beta}\delta g_{\alpha\beta} = 0\ \ \mathrm{on}\ \Gamma,\label{eq:bdry_condition}
\end{align}
where $\gamma_{\ \mu}^{\alpha} = \delta_{\ \mu}^{\alpha} - n^\alpha n_\mu $ is the projection tensor onto the boundary. With this choice of boundary condition it can be shown that
\begin{align}
n_\mu \omega^\mu = D_\mu \delta c^\mu\ \  \mathrm{on}\ \Gamma,
\end{align}
where $D_\mu$ is the hypersurface covariant derivative and
\begin{align}
c^\mu = \frac{1}{2\kappa} \gamma^{\mu}_{\ \alpha} \delta g^{\alpha\beta} n_\beta.\label{eq:c}
\end{align}
This follows after applying $\delta$ to equation (3.43) of \cite{Harlow}. The symplectic flux density is therefore a total derivative, and when integrated it generates contributions from the ``corners" where the Cauchy surfaces intersect~$\Gamma_{12}$. Therefore we obtain 
\begin{align}
0 = \Bigg(\int_{\Sigma_1} \mathrm{vol}\, t_\mu\omega^\mu + \int_{\d\Sigma_1} \mathrm{vol}\, \tau_\mu \delta c^\mu\Bigg) -\Bigg(\int_{\Sigma_2} \mathrm{vol}\, t_\mu\omega^\mu + \int_{\d\Sigma_2} \mathrm{vol}\, \tau_\mu \delta c^\mu \Bigg), 
\end{align}
which is an expression of conservation of the symplectic structure defined as 
\begin{align}
\Omega =\int_{\Sigma} \mathrm{vol}\, t_\mu\omega^\mu + \int_{\d\Sigma} \mathrm{vol}\, \tau_\mu \delta c^\mu.\label{eq:Cauchy_bdry_Omega}
\end{align}
The symplectic structure contains a contribution from the boundary of the Cauchy surface exactly as found in \cite{Harlow}.

However, there is another way we can try to get this result. We can simply compute the formula \eq{Omega} in the presence of the spatial boundary and see if it works. At first the computation goes in the same way as without boundary, giving \eq{GRPi}. However, now the improvement term makes a difference, creating a boundary correction to  \eq{GROmega} as
\begin{align}
\Omega = \int_M \vol\, \big(\nabla_\mu\sigma\big) \omega^\mu +\int_\Gamma \vol\, \big(D_\mu\sigma\big) B^{\mu\nu}n_\nu.
\end{align}
Presently we do not imagine any special choice of sigmoid. It can be nonzero on the spatial boundary but must satisfy \eq{sigmalimit} in the infinite past and future. The correct symplectic form for a generic sigmoid can be inferred from \eq{Cauchy_bdry_Omega} and takes the form
\begin{align}
\Omega = \int_M \vol\, \big(\nabla_\mu\sigma\big) \omega^\mu +\int_\Gamma \vol\, \big(D_\mu\sigma\big) \delta c^\mu. 
\end{align}
Therefore \eq{Omega} gives the correct symplectic structure if we can equate
\begin{align}
B^{\mu\nu}n_\nu = \delta c^\alpha\ \ \mathrm{on}\ \Gamma.
\end{align}
To see if this holds we must evaluate $\delta c^\alpha$. For this we need the differentials~\cite{Harlow},
\begin{subequations}
\begin{align}
\delta n^\mu &= \frac{1}{2}\big(\delta^\mu_{\ \alpha}+\gamma^\mu_{\ \alpha}\big) \delta g^{\alpha\beta}n_\beta\\
\delta n_\mu & = -\frac{1}{2}\big(g_{\mu\alpha}-\gamma_{\mu\alpha}\big)\delta g^{\alpha\beta} n_\beta.\label{eq:deltan_mu}
\end{align}
\end{subequations}
Comparing the first relation with \eq{c} we infer
\begin{align}
c^\mu = \frac{1}{\kappa} \delta n^\mu -\frac{1}{2\kappa} \delta g^{\mu \alpha}n_\alpha,
\end{align}
which implies that
\begin{align}
\delta c^\mu = \frac{1}{2\kappa}\delta g^{\mu\alpha} \delta n_\alpha.
\end{align}
Substituting \eq{deltan_mu} gives
\begin{align}
\delta c^\mu = -\frac{1}{4\kappa}\delta g^{\mu\alpha} \big(g_{\alpha\beta} -\gamma_{\alpha\beta}\big)\delta g^{\beta\nu}n_\nu.\label{eq:deltac1}
\end{align}
It is clear that $\delta c^\mu$ is orthogonal to $n_\mu$ by symmetry. This implies
\begin{align}
\delta c^\mu = \gamma^{\mu}_{\ \alpha}\delta c^\alpha
\end{align}
With the additional projection tensor the second term of \eq{deltac1} drops out on account of the boundary condition \eq{bdry_condition}. Therefore
\begin{align}
\delta c^\mu = -\frac{1}{4\kappa}\delta g^{\mu\lambda} g_{\lambda\rho}\delta g^{\rho\beta}n_\beta = B^{\mu\nu}n_\nu .
\end{align} 
This means that \eq{Omega} reproduces the symplectic structure of general relativity even in the presence of a finite spatial boundary. This is quite surprising. None of the arguments we have given for the consistency of \eq{Omega} generalize when spatial boundaries are present. The fact that the formula works anyway requires explanation. We leave this problem open to future work.

\subsection{$p$-adic string theory}
\label{subsec:padic}

As a final example we consider $p$-adic string theory \cite{Brekke}, a field theory model which has had some applications in the study of open string tachyon condensation \cite{Moeller,Ghoshal}. This theory is genuinely nonlocal. The symplectic structure on phase space is not widely known, but was recently obtained in a form which can be compared to ours in \cite{Heredia} using the formalism of \cite{Llosa,Gomis,Gomis2}. 

We consider $p$-adic string theory in one dimension (time), because its spatial nonlocality has a different character which for present discussion is not relevant. The action is
\begin{align} 
	S= {1 \over g^2} \int dt \left[ - {1 \over 2} \phi \,  p^{-{1 \over 2 }\Box} \phi  
	+ {1 \over p + 1 } \phi^{p + 1} 
	\right] \, ,
\end{align} 
where $\phi = \phi(t)$ is a real scalar field and $\Box=-\frac{d^2}{dt^2}$. The parameter $g$ is a coupling constant, and  $p$ is originally a prime number, but we only need $p>1$. The kinetic term contains an exponential of the d'Alembert operator which may be presented in a derivative expansion
\begin{align}
p^{-\frac{1}{2}\Box} = \sum_{n=0}^\infty \frac{1}{n!}\left(\frac{\ln p}{2}\frac{d^2}{dt^2}\right)^n.
\end{align}
In this sense the Lagrangian contains an infinite number of time derivatives. The derivative expansion however does not express the fundamental nonlocality of the kinetic term. This can be seen more clearly through its representation as a Weierstrass transform \cite{Moeller}
\begin{align}
p^{-\frac{1}{2}\Box} \phi(t) =  \frac{1}{\sqrt{2 \pi \ln p}} \int_{-\infty}^\infty d t' \, 
	\exp \bigg( -\frac{(t-t')^2}{2 \ln p} \bigg) \, \phi(t').
\end{align}
Nonlocality means that this operator probes the field far away from the point on which it acts.

The action for a linearized fluctuation $\varphi$ of the solution~$\phi$ is 
\begin{align}
S_\mathrm{fluctuation} = -\frac{1}{2g^2}\int dt\, \varphi \Big(p^{-\frac{1}{2}\Box} -p \phi^{p-1}\Big)\varphi.
\end{align}
From this we can read off the matrix elements $Q_{ij}$:
\begin{align}
Q(t,t') & = \frac{1}{g^2}\Big(p^{-\frac{1}{2}\Box} -p \phi^{p-1}\Big) \delta(t-t')\nonumber\\
& = \frac{1}{g^2\sqrt{2 \pi \ln p}} \exp \bigg( -\frac{(t-t')^2}{2 \ln p} \bigg)-p \phi^{p-1} \delta(t-t').
\end{align}
Note that $Q_{ij}$ is not a differential operator acting on a delta function. To make the symplectic form look as local as possible, we choose a step function sigmoid $\theta(t)$. The matrix $\sigma^i_j$ is then
\begin{align}
\sigma(t,t') = \theta(t)\delta(t-t').
\end{align}
The commutator of $Q_{ij}$ with $\sigma^i_j$ is
\begin{align}
\int_{-\infty}^\infty ds \Big[Q(t,s)\sigma(s,t')- \sigma(t,s)Q(s,t')\Big]= \frac{1}{g^2\sqrt{2 \pi \ln p}} \exp \bigg( -\frac{(t-t')^2}{2 \ln p} \bigg)\Big[\theta(t')-\theta(t)\Big].\label{eq:Qsigma_padic}
\end{align}
Substituting into \eq{Omegabasis} gives 
\begin{align}
\Omega = -\frac{1}{2}\frac{1}{g^2\sqrt{2 \pi \ln p}}\int_{-\infty}^\infty dt\int_{-\infty}^\infty dt' \,
\exp \bigg( -\frac{(t-t')^2}{2 \ln p} \bigg)\Big[\theta(t')-\theta(t)\Big]\delta\phi(t)\delta\phi(t').\label{eq:prepadicOmega}
\end{align}
Observe that $\theta(t)-\theta(t')$ vanishes when $t$ and $t'$ are positive. Therefore the integrand is nonzero only for the two quadrants where $t$ and $t'$ have opposite sign. The quadrants give the same contribution, resulting in 
\begin{align}
\Omega = -\frac{1}{g^2\sqrt{2 \pi \ln p}}\int_{-\infty}^0 dt\int_{0}^\infty dt' 
\exp \bigg( -\frac{(t-t')^2}{2 \ln p} \bigg)\delta\phi(t)\delta\phi(t').\label{eq:padicOmega}
\end{align}
The symplectic structure is not a function of the field and its derivatives on a time slice.  Instead, it correlates the field in the future of $t=0$ to the field in the past to a degree which is suppressed by a Gaussian of their separation in time. This makes it appear as though fields in the future and past of $t=0$ are conjugate phase space variables. But this cannot be right because the symplectic pairing on phase space cannot depend on the time slice. What this tells us is that fields at different times are not independent phase space variables in $p$-adic string theory. The equation of motion implies an infinite set of constraints which relate them. This is different from nonlocal theories of the kind discussed in~\cite{Woodard}, where fields at different times can be independent. It is not known how to find complete and independent degrees of freedom on the phase space of $p$-adic string theory. Some aspects of this problem are discussed in \cite{Moeller,Barnaby}.

Let us understand why the $p$-adic symplectic form is conserved. We can shift the ``time slice" to $t=t_0$ by replacing the upper/lower limits of the two integrals in \eq{padicOmega} with $t_0$. Evaluating the derivative with respect to $t_0$ gives
\begin{align}
	\frac{d}{dt_0}\Omega = -\frac{1}{g^2\sqrt{2 \pi \ln p}}\Bigg[&\int_{t_0}^\infty dt' 
	\exp \bigg( -\frac{(t_0-t')^2}{2 \ln p} \bigg)\delta\phi(t_0)\delta\phi(t')\nonumber\\
	& -\int_{-\infty}^{t_0} dt 
	\exp \bigg( -\frac{(t-t_0)^2}{2 \ln p} \bigg)\delta\phi(t)\delta\phi(t_0)\Bigg].
\end{align}
Switching the order of the $\delta\phi$'s and relabeling the integration variables the two terms can be combined into a single integral:
\begin{align}
	\frac{d}{dt_0}\Omega = -\frac{1}{g^2\sqrt{2 \pi \ln p}}\delta\phi(t_0) \int_{-\infty}^\infty dt 
	\exp \bigg( -\frac{(t -t_0)^2}{2 \ln p} \bigg)\delta\phi(t).
\end{align}
Recognizing the Weierstrass transform we can write this as 
\begin{align}
	\frac{d}{dt_0}\Omega = -\frac{1}{g^2}\delta\phi(t_0)\Big( p^{\frac{1}{2}\frac{d^2}{dt_0^2}}\delta\phi(t_0)\Big).
\end{align}
Because the exterior derivative on pre-phase space must preserve the equations of motion, $\delta\phi$ satisfies 
\begin{align}
	\Big(p^{-\frac{1}{2}\Box} -p \phi^{p-1}\Big)\delta\phi= 0.\label{eq:QdeltaPhi_padic}
\end{align}
This implies
\begin{align}
	\frac{d}{dt_0}\Omega = -\frac{1}{g^2}\delta\phi(t_0)\Big( p \phi^{p-1}\delta\phi(t_0)\Big)=0,
\end{align}
because $\delta\phi$ squares to zero. 

It is interesting to understand why the symplectic structure does not vanish. Earlier discussion phrased this as an issue of total derivatives and boundary terms, but this doesn't seem adequate to explain the present example because there are no derivatives anywhere, at least when the kinetic operator is expressed through the Weierstrass transform. Let us first understand the sense in which the $p$-adic symplectic form might be thought to vanish. Recall from \eq{Qsigma_padic} that the commutator of $Q_{ij}$ and $\sigma^i_j$ created a difference of step functions. If this difference is broken into two terms which are treated separately, the $p$-adic symplectic structure formally appears as
\begin{align}
	\Omega = -\frac{1}{2}\frac{1}{g^2\sqrt{2 \pi \ln p}}\Bigg[&\int_{-\infty}^\infty dt\int_{0}^\infty dt' 
	\exp \bigg( -\frac{(t-t')^2}{2 \ln p} \bigg)\delta\phi(t)\delta\phi(t')\nonumber\\
	&-\int_{0}^\infty dt\int_{-\infty}^\infty dt' 
	\exp \bigg( -\frac{(t-t')^2}{2 \ln p} \bigg)\delta\phi(t)\delta\phi(t')\Bigg].\label{eq:padicOmega_break}
\end{align}
In the first term we recognize the Weierstrass transform in the integral over $t$ and the second term in the integral over $t'$. This allows us to simplify
\begin{align}
	\Omega = -\frac{1}{2g^2}\Bigg[\int_{0}^\infty dt\, \Big(p^{-\frac{1}{2}\Box}\delta\phi(t)\Big) \delta\phi(t)-\int_{0}^\infty dt\, \delta\phi(t)\Big(p^{-\frac{1}{2}\Box}\delta\phi(t)\Big)\Bigg].\label{eq:padicOmega_zero}
\end{align}
Using \eq{QdeltaPhi_padic} we find zero by $(\delta\phi)^2=0$. As might be expected, the difficulty arises from separating the difference of step functions into two parts. Each part contributes a double integral to the final result which is not absolutely convergent. The ambiguity comes from the quadrant where $t$ and $t'$ are positive, exactly the quadrant which was argued to cancel below \eq{prepadicOmega}. This quadrant contains the line $t=t'$ where the Gaussian does not suppress the integration towards infinity. The resolution, as we have seen in other examples, is to apply the tau regulator to resolve ambiguities.  Let us replace $\delta\phi$ with $\tau\delta\phi$ where $\tau$ is a step function which implements a cut off on the limits of integration over time. Factoring out the integrand, \eq{padicOmega_break} appears as
\begin{align}
	\Omega = -\frac{1}{2}\frac{1}{g^2\sqrt{2 \pi \ln p}}\lim_{T\to\infty}\Bigg[\int_{-T}^T dt\int_{0}^T dt' -\int_{0}^T dt\int_{-T}^T dt' \Bigg] 
	\exp \bigg( -\frac{(t-t')^2}{2 \ln p} \bigg)\delta\phi(t)\delta\phi(t').\label{eq:padicOmegatau}
\end{align}
It is easy to see that integration over the positive $t,t'$ quadrant cancels between the two double integrals, leaving the correct nonzero answer~\eq{padicOmega}. The incorrect zero answer would have resulted if the symplectic form had been regularized as 
\begin{align}
	-\frac{1}{2}\frac{1}{g^2\sqrt{2 \pi \ln p}}\lim_{T\to\infty}\lim_{U\to\infty}\Bigg[\int_{-U}^U dt\int_{0}^T dt' -\int_{0}^T dt\int_{-U}^U dt' \Bigg] 
	\exp \bigg( -\frac{(t-t')^2}{2 \ln p} \bigg)\delta\phi(t)\delta\phi(t'),\label{eq:padicOmegatau}
\end{align}
where $U$ and $T$ are both large but $U$ is much larger than $T$. The difference between this and the correct expression is integration between $T$ and $U$ in the extreme future. In a few short steps one can show that this contributes as
\begin{align}
	\Omega = -\frac{1}{g^2\sqrt{2 \pi \ln p}}\lim_{T\to\infty}\int_{-\infty}^T dt\int_{T}^\infty dt' 
	\exp \bigg( -\frac{(t-t')^2}{2 \ln p} \bigg)\delta\phi(t)\delta\phi(t'), 
\end{align}
which is the symplectic form itself with the time slice pushed to infinity. 

\begin{figure}[t]
	\centering
	\includegraphics[scale=1.75]{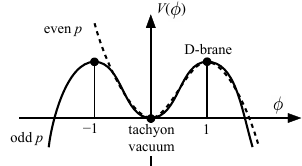}
	\caption{\label{fig:CovL4} Scalar potential in $p$-adic string theory. For both even and odd $p$ the potentials are unbounded from below.} 
\end{figure} 

We want to test the symplectic structure with a physical application. In a similar spirit as Cho, Mazel and Yin \cite{Cho}, we will use it to compute the energy of decaying D-branes. The potential of $p$-adic string theory is given by setting all derivatives in the Lagrangian to zero:
\begin{align}
	V(\phi) = {1 \over g^2} \left(
	{1 \over 2} \phi^2 - {1 \over p+1} \, \phi^{p+1}
	\right).
\end{align}
There is always a stable vacuum at $\phi = 0$ and an unstable vacuum at $\phi=1$. For odd $p$ there is an additional unstable vacuum at $\phi = -1$. See figure \ref{fig:CovL4}. The unstable vacuum $\phi=1$ represents a D-brane in $p$-adic string theory. The stable vacuum $\phi=0$ represents the final state after the D-brane has disappeared, and is called the {\it tachyon vacuum}. The linear wave equation for a fluctuation of the D-brane is
\begin{align}
 \Big(p^{-{1 \over 2 }\Box} - p\Big)\varphi  = 0 .
\end{align}
The simplest solutions are exponentials which set $-\frac{1}{2} \Box=1$. The exponential growth represents the disintegration of the D-brane. We consider solutions which have lower energy than the D-brane itself, so the field rolls up the potential but falls short of the maximum before turning around. The linearized solution of this kind is
\begin{align}
\varphi = -\lambda \cosh \sqrt{2}(t-t_0). \label{eq:cosinepadic}
\end{align}
This can be promoted to a full nonlinear solution by adding corrections as a power series in~$\lambda$. The result is a $p$-adic analogue of the rolling tachyon solutions discussed by Sen \cite{Sen}. Our goal is to find the energy of these solutions as a function of $\lambda$. This is nontrivial already at leading order because the $p$-adic kinetic term is nonlocal. If we had the standard Klein-Gordon kinetic term the energy would be given by the depth of the potential at the turnaround point, where the velocity of the field vanishes. But the energy in $p$-adic string theory receives contribution not only from velocities, but from accelerations and all higher order time derivatives of the field. The energy in $p$-adic string theory has been determined via the Noether procedure  by Moeller and Zwiebach \cite{Moeller}, who give the result as a derivative expansion
\begin{align}
	E = {1 \over 2 g^2 } {p-1 \over p+1} \phi^{p+1} 
	- {1 \over 2 g^2} \sum_{l=1}^\infty {1 \over l!} \left({1 \over 2} \ln p\right)^l
	\sum_{m=0}^{2l -1} (-1)^m \left.\frac{d^m\phi}{dt^m}\frac{d^{2l-m}\phi}{dt^{2l-m}}\right|_{t=t_0}.
\end{align}
Substituting \eq{cosinepadic} gives energy of the decaying D-brane up to leading nontrivial order 
\begin{align}
	E(\lambda) = \left({1 \over 2 g^2 } {p-1 \over p+1}\right) - \left({p \ln p \over 2 g^2}\right) \lambda^2\,  + \mathcal{O}(\lambda^3).\label{eq:NoetherE}
\end{align}
For comparison, if we had the $p$-adic potential but the standard Klein-Gordon kinetic term, the coefficient at order $\lambda^2$ would be $-\frac{p-1}{2g^2}$. 

We now attempt to verify this result using the symplectic form. Presently we have a two-dimensional phase space of solutions, parameterized by $\lambda$, which represents the displacement of the field from the maximum at the turnaround point, and $t_0$, which represents the time when the turnaround occurs. Evaluating the symplectic form on this phase space will give an expression
\begin{align}
\Omega = \Omega(\lambda) \delta t_0\, \delta\lambda,
\end{align}
where $\Omega(\lambda)$ is a number-valued function of $\lambda$ only; it will not depend on $t_0$ because the symplectic form is conserved. We may further express this as
\begin{align}
\Omega = \delta t_0 \delta E(\lambda),
\end{align}
where $E(\lambda)$ should be identified with the energy of the time-dependent decay process. This is the quantity we will compare to the Noether energy computed in \eq{NoetherE}. Since we want the energy at leading nontrivial order it is sufficient to compute the differential of the linearized solution~\eq{cosinepadic}
\begin{align}
\delta\varphi = \Big(-\cosh\sqrt{2}(t-t_0)\Big)\delta\lambda +\Big(-\sqrt{2}\lambda\sinh\sqrt{2}(t-t_0)\Big)\delta t_0,
\end{align}
and plug into \eq{padicOmega}. Making use of conservation we set $t_0$ to zero (but $\delta t_0$ is still nonzero). This produces an integral
\begin{align}
\Omega = \delta t_0 \delta\lambda \left(\frac{2\sqrt{2}\lambda }{g^2\sqrt{2 \pi \ln p}}\int_{-\infty}^0 dt\int_{0}^\infty dt' 
\exp \bigg( -\frac{(t-t')^2}{2 \ln p} \bigg)\sinh \sqrt{2}t\cosh\sqrt{2} t'\right) + \mathcal{O}(\lambda^2).
\end{align}
Renaming $t\to-t$ and using an ``angle-sum'' formula simplifies to
\begin{align}\Omega & = 
-\delta t_0 \delta\lambda \left(\frac{\sqrt{2}\lambda }{g^2\sqrt{2 \pi \ln p}}\int_{0}^\infty dt \int_{0}^\infty dt' 
\exp \bigg( -\frac{(t+t')^2}{2 \ln p} \bigg)\sinh \sqrt{2}(t+t') \right) + \mathcal{O}(\lambda^2)\nonumber\\
& = -\delta t_0 \delta\lambda \left(\frac{\sqrt{2}\lambda }{g^2\sqrt{2 \pi \ln p}}\int_{0}^\infty d u\, u \sinh\big(\sqrt{2}u\big)\, e^{-u^2/(2\ln p)}\right) + \mathcal{O}(\lambda^2).
\end{align}
The remaining expression can be computed by extending the range of integration down to $-\infty$, replacing the $\sinh$ with an exponential, and evaluating the Gaussian integral. The result is
\begin{align}
\Omega= -\delta t_0\delta \lambda \frac{\lambda p\ln p}{g^2} + \mathcal{O}(\lambda^2).
\end{align}
This can be rewritten 
\begin{align}
\Omega & = \delta t_0\,\delta\Bigg[\left({1 \over 2 g^2 } {p-1 \over p+1}\right) - \left({p \ln p \over 2 g^2}\right) \lambda^2\,  + \mathcal{O}(\lambda^3)\Bigg] \, ,
\end{align}
which is in agreement with the Noether energy in \eq{NoetherE}. Indeed, the symplectic form is not zero. 

\section{Future directions}
\label{sec:prospects}

We list some directions for future work: 
\begin{itemize}
\item A satisfactory version of the Hamiltonian formalism requires a Poisson bracket and conserved charges. These objects have a very elegant description in terms of $L_\infty$ algebras. We hope to describe the results in upcoming work~\cite{Bernardes}, but for the sake of interest it is worth seeing the Hamiltonian. The expression is reminiscent of the action \eq{action},
\begin{align}
H = \frac{1}{2}\omega( \Phi,H_1\Phi)+\sum_{n=2}^\infty \frac{1}{(n+1)!}\omega\big(\Phi,H_n(\underbrace{\Phi,\cdots,\Phi}_{n\ \mathrm{times}})\big),\label{eq:Hamiltonian}
\end{align}
but the products $H_n$ are different. They are related to the products $L_n$ of the action through 
\begin{align}
H_n(\Phi,\cdots,\Phi) = iE \sigma L_n(\Phi,\cdots,\Phi) -n L_n(\sigma iE \Phi,\Phi,\cdots,\Phi),
\end{align}
where $iE$ is the operator generating time translations in $\H$ (the time derivative) and $\sigma$ is the sigmoid. The sigmoid plays a similar role for the Hamiltonian as it does for the symplectic structure. It is striking how different this formula appears from the standard Legendre transform. The most significant feature is the absence of explicit canonical momenta, whose construction is the main source of difficulty in the Hamiltonian description of nonlocal field theories.

\item We have not discussed how the symplectic structure is related to the variational problem in Lagrangian mechanics. This is important for understanding spatial boundaries and associated corrections to observables such as the Hamiltonian and action \cite{Harlow,Arnowitt,York,Gibbons}. This is connected to the question of why \eq{Omega} computes the symplectic general relativity even with finite spatial boundary. 

\item The clearest advantage of the formula \eq{Omega} is in dealing with nonlocality. Nonlocal field theories have been studied  as a mechanism to remove ultraviolet divergences in quantum field theory~\cite{Efimov,Tomboulis,Modesto} and for applications to cosmology \cite{Arefeva,Biswas,Deser}. Our primary motivation has been to give a viable Hamiltonian description of string field theory \cite{Erbin,Sen2}, which could open the path to many new calculations within string theory.  Examples include conserved quantities in time-dependent string backgrounds \cite{Cho}; D-brane charges~\cite{Townsend} and gravitational mass; black hole entropy~\cite{Wald,Susskind,Ahmadain,Halder} or entanglement entropy~\cite{Balasubramanian,Hubeny,Donnelly,Naseer}. In upcoming work we hope to establish that the symplectic structure \eq{Omega} is well-defined and computable in open string field theory \cite{Bernardes2}. The generalization to closed string field theory is important for addressing more subtle quantum gravity questions. Diffeomorphism invariance~\cite{Mazel,Mamade1,Mamade2} implies that spacetime boundaries must play a significant role in extracting observables \cite{Erler2,Stettinger,Firat,Maccaferri1,Maccaferri2}.  

\item Once we have phase space it is natural to think about quantization. It seems likely that loop homotopy algebras \cite{Zwiebach,Markl} will be a critical part of extending the formalism to the quantum level. It is interesting to think about how operator ordering ambiguities might be resolved in the context of string theory, where quantization is expected to be unique. 
\end{itemize}

\subsection*{Acknowledgments}

The authors thank  M. Cho, H. Erbin, D. Ghoshal, D. Gross, I. Khavkine, I. Kola{\v r}, A. Maharana, M. Rangamani, A. Sen, A. Vikman, X. Yin and B. Zwiebach for discussions.  This work was initiated during the program ``What is String Theory? Weaving Perspectives Together'' at Kavli Institute for Theoretical Physics (KITP) in Santa Barbara, and in the workshop  ``Matrix models and String field theory'' at the Pedro Pasual Center for Sciences in Benasque, Spain. We thank D. Gross and the KITP for hospitality while carrying out part of this work. The work of VB and TE was supported by the European Structural and Investment Funds and the Czech Ministry of Education, Youth and Sports (project No. FORTE—CZ.02.01.01/00/22\_008/0004632) and the work of TE was supported in part by grant NSF PHY-2309135 to the KITP. The work of AHF is supported by the U.S. Department of Energy, Office of Science, Office of High Energy Physics of U.S. Department of Energy under grant Contract Number DE-SC0012567, DE-SC0009999, and the funds from the University of California.

\end{document}